\documentclass[aps,prl,twocolumn,showpacs,unsortedaddress]{revtex4}

\usepackage[dvips]{graphicx}
\usepackage{amsmath}
\usepackage{amsfonts}
\usepackage{amssymb}
\usepackage[latin1]{inputenc}
\usepackage{xcolor}
\usepackage{subfigure}

\definecolor{darkblue}{rgb}{0, 0, 0.8}
\usepackage[colorlinks=true, breaklinks=true, linkcolor=darkblue, citecolor=darkblue, urlcolor=darkblue]{hyperref}
\newcommand{\doilink}[2]{\href{http://dx.doi.org/#1}{#2}}

\begin{document}

\title{Propagation of light through small clouds of cold interacting atoms}

\author{S.~Jennewein, Y.R.P.~Sortais, J.-J.~Greffet and A.~Browaeys}
\affiliation{Laboratoire Charles Fabry, Institut d'Optique, CNRS, Univ Paris Sud,
2 Avenue Augustin Fresnel,
91127 Palaiseau cedex, France}

\date{\today}

\begin{abstract}

We demonstrate experimentally that a cloud of cold atoms with a size comparable to the wavelength of light can induce large group delays on a laser pulse when the laser is tightly focused on it and is close to an atomic resonance. Delays as large as $-10$\,ns are observed, corresponding to ``superluminal" propagation with negative group velocities as low as $-300$\,m/s. Strikingly, this large delay is associated with a moderate extinction owing to the very small size of the cloud and to the light-induced interactions between atoms. It implies that a large phase shift is imprinted on the continuous laser beam, and opens interesting perspectives for applications to quantum technologies.
\end{abstract}

\maketitle
Resonant media are able to sustain the propagation of wave packets with variable group velocities, depending on the frequency detuning of the wave with respect to the resonance frequency. This property can lead to slow light, associated to large temporal delays in the transmission of a pulse of light, or even light with negative group velocities (see~\cite{Withayachumnankul2010} for a recent review). The control of the group velocity $v_{\rm g}$ relies on the strong variation, close to resonance, of the refractive index $n(\omega)$ of the medium with the frequency of light, leading to a group index $n_g= c/v_{\rm g}=n+\omega~dn/d\omega$ ($c$ is the speed of light) that can be smaller than unity or negative~\cite{Zangwill2012}. The latter situation seems paradoxical (the pulse of light seems to exit the medium before it entered it) and corresponds to the so-called super-luminal propagation~\cite{Garrett1970,Gehring2006}. In any case, situations where the group index is large enough to lead to a significant delay are associated, through the Kramers-Kronig relations~\cite{Zangwill2012}, with a strong attenuation of the light, making the delay hard to measure. To circumvent this problem, controlling the group velocity in an optical atomic medium often relies on electromagnetically induced transparency (EIT)~\cite{Fleischhauer2005}. This dispersion management has been used with ultra-cold atomic clouds to produce light with velocities as low as $17$\,m/s~\cite{Hau1999}. In parallel, a few experiments have used a single resonance to control the group velocity inside passive media such as a thin sample of GaP:N~\cite{Chu1982}, a centimeter-long sample of room temperature atomic vapor~\cite{Tanaka2003} or a sub-micrometer thick slab of hot atomic vapor~\cite{Keaveney2012b}.

The group delay $\tau(\omega)$ is also interesting as it is related to the phase $\phi(\omega)$ imprinted by the medium on the light propagating through it : $\tau(\omega)=d\phi/d\omega$~\footnote{The group delay $\tau(\omega)=\phi'(\omega)$ should not be confused with the Wigner delay~\cite{Wigner1955,Bourgain2013a,Jechow2013}, which has the same expression, but with $\phi$ the phase of the atomic dipoles with respect to the driving field.}. As such the measurement of the group delay provides a measurement of the phase. Quantum technologies would benefit from the capability to imprint a large phase shift on an individual photon, possibly switchable by another individual photon~\cite{Chang2014}. Impressive results have been obtained using individual ions~\cite{Jechow2013}, atoms~\cite{Abdullah2009}, or molecules~\cite{Pototschnig2011} to imprint phase shifts of a few degrees on a tightly focused laser beam in free space. A recent experiment also demonstrated a phase shift of $18\,\mu$rad on a post-selected photon passing through a large, dilute cloud of cold atoms~\cite{Feizpour2015}. But larger phase shifts would be desirable, and using small atomic ensembles containing more than one atom could provide them.

Here, we show that a microscopic cloud containing a few hundred cold atoms placed at the focal point of a tightly focused laser beam can induce large group delays when tuned and probed near a single atomic resonance. Our situation is special in that the size of the atomic medium is much smaller than the size of the beam, and yet it induces large delays with an only moderate extinction. Experimentally, we send a pulse with a weak intensity (less than one photon on average) and a Gaussian temporal envelope onto the cloud. We measure delays as large as $10$\,ns, with values either positive or negative depending on the detuning of the laser with respect to the atomic resonance. The corresponding group velocities can be either negative or positive with values as low as $300$\,m/s. Strikingly, we observe a fractional pulse advancement as large as $40\%$ of the pulse width with a transmission not smaller than $20\%$. The observed delays are in good agreement with the independent measurement of the steady-state coherent optical response of the cloud that we performed recently~\cite{Jennewein2015a}. Our measurements imply that when sending a weak continuous laser beam on the cloud, phase shifts as large as $\sim 1$\,rad are imprinted on it, making it potentially interesting for applications in quantum technologies.

\begin{figure}
\includegraphics[width=\columnwidth]{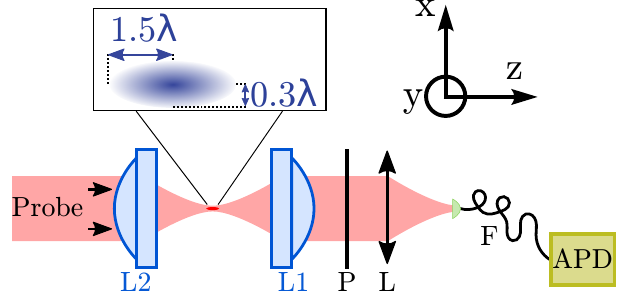}
\caption{Experimental setup. A microscopic cloud of $^{87}$Rb atoms is illuminated by a linearly-polarized probe laser focused down to a waist $w=1.2~\mu$m. P: polarization beam-splitter, aligned at $45^\circ$ of the laser polarization. L: lens allowing the mode-matching between the probe beam and the fiber mode in the absence of atoms. F: single-mode fiber. APD: avalanche photodiode.}
\label{fig1_setup}
\end{figure}

Our setup is represented in Fig.~\ref{fig1_setup}. It consists of two identical aspherical lenses with a high numerical aperture (NA=$0.5$) mounted in a confocal configuration in a vacuum chamber~\cite{Sortais2007}. The lens L1 is used to focus a laser beam at a wavelength of $940$\,nm onto a waist of $1.2\,\mu$m, thus creating a dipole trap with a depth of $1$\,mK. We load the trap with $^{87}$Rb atoms, which have a temperature of $120\,\mu$K, resulting in a cigar-shaped cloud with transverse and longitudinal root-mean-square (RMS) widths $L_\perp=0.2\,\mu$m and $L_z=1.2\,\mu$m respectively~\cite{Bourgain2013b}. We control the number of atoms $N$ within $10\%$, and can vary $N$ between $10$ and $180$~\footnote{The exact number of atoms inside the trap varies shot to shot as it is governed by a sub-Poisson distribution with a mean $N$ and variance $3N/4$~\cite{Sortais2012}.}. The probe beam is focused by the lens L2 at the position of the cloud down to a waist $w=1.20\pm0.05\,\mu$m, larger than the cloud transverse size. The probe light is linearly polarized and is nearly resonant with the D2 closed transition of rubidium between the $(5S_{1/2},F=2)$ and $(5P_{3/2},F=3)$ levels at $\lambda =2\pi c/\omega_0= 780.2$\,nm (linewidth $\Gamma = 2\pi\times 6$\,MHz)~\footnote{The magnetic field is compensated to values lower than $\sim 80$\,mG.}. The intensity of the probe is $I/I_{\rm sat}=0.04$ ($I_{\rm sat}=1.6$\,mW/${\rm cm}^2$), small enough to be in the elastic scattering regime. The probe light transmitted through the cloud is collected using L1 and coupled into a single-mode fiber connected to an avalanche photodiode. The temporal signals are acquired by accumulating single photons using a counting card with a resolution of $150$\,ps.

Our experimental arrangement measures the interference of the incoming probe light ${\bf E}_{\rm L}$ and the field coherently scattered by the cloud ${\bf E}_{\rm sc}$ (see~\cite{Jennewein2015a} for more details), leading to a total field ${\bf E}={\bf E}_{\rm L}+{\bf E}_{\rm sc}$. The use of a single-mode fiber implies that, in steady state, we measure the overlap between ${\bf E}$ and the fiber mode ${\bf g}$, ${\cal E}(\omega)= \int \left\{{\bf E}({\bf r},\omega)\cdot {\bf g}^*({\bf r})\right\}dS$ ($dS$ is a differential area element perpendicular to the optical axis)~\cite{Abdullah2009,Tey2009}. The setup is aligned such that the fiber mode is matched to the incoming light, i.e. ${\bf g}\propto {\bf E}_{\rm L}$. As detailed in~\cite{Jennewein2015a}, we thus define the coherent optical response of the cloud in steady state by a transfer function ${\cal S}(\omega)={\cal E}(\omega)/{\cal E}_{\rm L}(\omega)$ that links the detected fields with and without atoms. The group delay is then related  to the phase of the transfer function $\phi(\omega)=\arg[{\cal S}(\omega)]$ by $\tau(\omega)=\phi'(\omega)$~\cite{Zangwill2012}. Measuring the group delay provides a way to access the phase of the transfer function, i.e. the phase imprinted by the cloud on the laser beam, and is an alternative to direct, e.g. interferometric, measurements of the phase~\cite{Abdullah2009,Pototschnig2011}.

\begin{figure}
\includegraphics[width=\columnwidth]{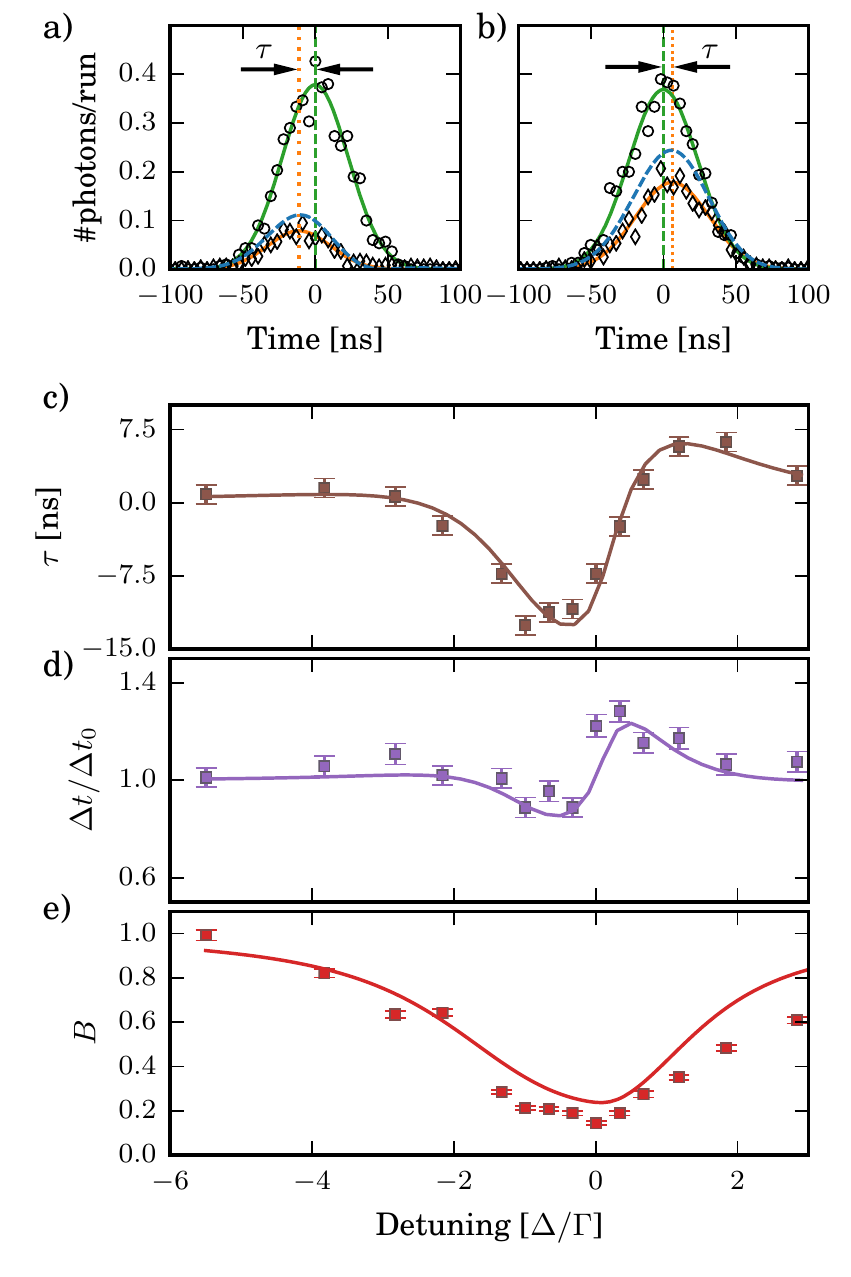}
\caption{Temporal signals recorded on the APD. Circles: Gaussian reference pulse (no atoms). Lozenges: transmitted pulse
for  $N=170$. (a) $\Delta=-0.7\Gamma$ : the pulse is maximally advanced by $10$\,ns. (b) $\Delta=1.8\Gamma$ : the pulse is maximally retarded by $7$\,ns. Solid lines: Gaussian fits to the data. Dashed lines: reconstructed pulse using Eq.~\ref{Eq_Fouriertransform}. Each run
consists of $1000$ illuminations with a duration of $300$\,ns each. (c) Group delay $\tau$. (d) RMS duration $\Delta t$ of the pulse detected in the presence of atoms, normalized to the case without atoms. (e) Amplitude $B$ of the transmitted pulse. In (c), (d), (e), the error bars are from the Gaussian fit of the detected pulse, and the solid lines are the Gaussian fit parameters of the reconstructed pulse using Eq.~\ref{Eq_Fouriertransform}.}
\label{Fig2_pulse_delay}
\end{figure}

In order to measure the group delay, we proceed in the following way. After preparing the atoms in the  $(5S_{1/2},F=2)$ level, we alternately illuminate the cloud $1000$ times with the trap light and the probe light with a period of $1\,\mu$s. The probe light is sent during the $500$\,ns switch-off period of the dipole trap. By shaping the amplitude of the radio-frequency input signal of an acousto-optics modulator (AOM) we produce a probe pulse with a Gaussian temporal profile. The RMS duration of the intensity profile is $\Delta t_0 = 25$\,ns, leading to a bandwidth smaller than the bandwidth $\Gamma_c$ of the medium, which is  in the range $1.5\Gamma-3\Gamma$ for $10\lesssim N\lesssim 170$~\cite{Jennewein2015a}. The parameters of the probe imply that each pulse contains only $\sim 0.4$ photons.
We repeat the probe-trap-probe alternated illumination $200$ times with a new atomic cloud each time and the whole sequence for various probe detunings $\Delta = \omega - \omega_0$ and various atom numbers.

Figures~\ref{Fig2_pulse_delay}(a,b) show two examples of transmitted pulses, together with the pulse in the absence of atoms. Figure~\ref{Fig2_pulse_delay}(a) corresponds to a situation of an advanced  pulse and Fig.~\ref{Fig2_pulse_delay}(b) to the one of a retarded pulse. We find that the transmitted pulses are well fitted by a Gaussian function $G(t) = B \exp[-(t-\tau)^2/ 2 \Delta t^2]$ whatever the atom number and the detuning. The variations of the width $\Delta t$ and delay $\tau$ of the transmitted pulse are shown in Figs.~\ref{Fig2_pulse_delay}(c,d) as a function of the detuning $\Delta$ of the probe laser, for $N=170$ atoms~\footnote{The results of the fits for other values of the atom number used in this work are presented in the Supplemental Material.}. Close to resonance, we observe large negative group delays, corresponding to a ``superluminal" pulse propagation~\footnote{The situation where $v_{\rm g} > c$  corresponds to advanced delays $\sim L_z (1/c-1/v_{\rm g})<L_z/c\sim 1$\,fs, completely negligible to be observed in our experiment. We are therefore mainly sensitive to the regime where the group velocity is negative.}. The largest measured advanced delay is $-10$~ns, corresponding to a fractional pulse advancement $\tau/\Delta t \simeq 40\%$ with a transmission of $20\%$.
For comparison, Ref.~\cite{Keaveney2012b} reports a fractional pulse advancement of $55\%$, but with only $1\%$ transmission. We also observe pulses retarded by as much as $7$\,ns.

Considering that the cloud has a longitudinal size $L=\sqrt{2\pi} L_z$, an advanced  delay $\tau=-10$\,ns corresponds to a group velocity $v_{\rm g}= -300$\,m/s. As the cloud is much smaller than the beam waist, the link to a group index is awkward, but what the group index of a slab of thickness $L$ would be remains a valid question. The corresponding group index would be $|n_{\rm g}|= 10^6$, probably one of the largest ever measured on a single resonance. This large value is all the more remarkable that it occurs close to an atomic resonance with as much as $20\%$ transmission. The associated limited extinction is due to the small size of the cloud and to the fact that it is very dense: the light-induced dipole-dipole interactions lead to a saturation of the absorption with increasing number of atoms, as we have shown in~\cite{Jennewein2015a}.
We also note that such a large group index would correspond to a pulse compression by a factor $1/|n_{\rm g}|$~\cite{Fleischhauer2005}, giving for our $25$-ns long pulse a compression from $7.5$\,m to $10\,\mu$m. Despite the large compression factor, this value is larger than the cloud longitudinal size, indicating that the strong dispersion is not sufficient to entirely store the pulse inside the cloud, which  rather acts as an efficient delay line.

In order to understand quantitatively the measured delays and their variations with the laser detuning, we check that our results are consistent with our recent, independent measurement of the coherent transfer function ${\cal S}(\omega)$ of the cloud~\cite{Jennewein2015a}. Briefly, this function was measured in steady-state by using the exact same sequence, but having replaced the Gaussian temporal pulses by flat-top pulses with duration $300$\,ns (rise time $2$\,ns). We monitored the amount of transmitted light and waited until the signal reaches steady-state. We then divided the signal obtained in the presence of atoms with the one obtained without atoms, which is equal to the quantity $|{\cal S}(\omega)|^2$. We repeated this procedure for various detunings and atom numbers. We then fitted the data, using for the transfer function the following functional form:
\begin{equation}\label{Eq:S(omega)}
{\cal S}(\Delta,N) = 1-\frac{A}{1-2i\frac{\Delta-\Delta_c}{\Gamma_c}} \ ,
\end{equation}
with $A(N)$, $\Delta_{\rm c}(N)$ and $\Gamma_{\rm c}(N)$ as free parameters for a given atom number $N$
(see Supplemental Material). As shown in~\cite{Jennewein2015a},  for a cloud with a size much smaller than $\lambda/(2\pi)=1/k$, the amplitude $A$ would scale as $1/(kw)^2$, $w$ being the probe waist at the position of the atoms: this comes from the fact that a polarizable, non-lossy particle illuminated by a radiation of wavelength $\lambda$ has an optical cross section $6\pi/k^2$ at resonance, whatever its real physical size as soon as it is smaller than $1/k$. Although our cloud is not exactly in this regime, this suggests that a tight focus leads to the largest effect of the cloud on the probe laser~\cite{Abdullah2009,Tey2009}. In particular, the phase shift increases when the waist becomes small. In our case the phase shift can be as large as $\Delta\phi \sim 1$\,rad (see below). As the width $\Gamma_{\rm c}$ of the resonance, although broadened by the interactions, remains very small with respect to the resonance frequency, the group delay $\tau\sim \Delta\phi/ \Gamma_{\rm c}$ can reach tens of nanoseconds.

The frequency variations of the group delay presented in Fig.~\ref{Fig2_pulse_delay} show an asymmetric behaviour near resonance. This asymmetry is surprising since the functional form~(\ref{Eq:S(omega)}) indicates that the phase $\phi(\omega)$ has a dispersive behaviour, and thus $\tau(\omega)=\phi'(\omega)$ should be symmetric around the resonance. To understand the observed asymmetry we simulate the transmitted pulse intensity using the complex transfer function ${\cal S}(\omega)$ measured in steady state, and the following relationship:
\begin{equation}\label{Eq_Fouriertransform}
I(t)=\left|\int_{-\infty}^{\infty}  {\cal S}(\omega) {\cal E}_{\rm L}(\omega) e^{-i\omega t}\frac{d\omega}{2\pi}\, \right|^2 \ ,
\end{equation}
where we introduce a small linear frequency chirp $\beta$ in the spectral amplitude of the laser field :
\begin{equation}\label{Eq_Electric_Field}
{\cal E}_{\rm L}(\omega) = \frac{\sigma_{\cal E}}{\sqrt{1+2 i\beta\sigma_{\cal E}^2}}~e^{-(\omega-\omega_{\rm L})^2\sigma_{\cal E}^2/2(1+2 i \beta \sigma_{\cal E}^2)}.
\end{equation}
Here, $\omega_{\rm L}$ is the central frequency of the laser, and $\sigma_{\cal E} = \sqrt{2} \Delta t_0$. The introduction of the frequency chirp is motivated by our finding of a distorted frequency response of the AOM driver on the short time scale used here. We  measured $\beta =37$\,kHz/ns directly on the light. We fit the calculated pulse by a Gaussian function with the delay, width and amplitude as free parameters. Figure~\ref{Fig2_pulse_delay} shows a good agreement between the data measured directly on the transmitted pulse and the data extracted from the reconstructed pulses. This feature is verified for all the numbers of atoms used in this work (see Supplemental Information). In particular, the pulse reconstruction reproduces the measured variations of the width $\Delta t$ across the resonance, which are due to the dispersion $\phi''(\omega)$ and the pulse spectral width being not very small with respect to the atomic resonance linewidth $\Gamma_{\rm c}$. We emphasize that the asymmetry observed in the delay and width variations comes from the chirp on the laser frequency, and not from the response of the medium.

\begin{figure}
\includegraphics[width=\columnwidth]{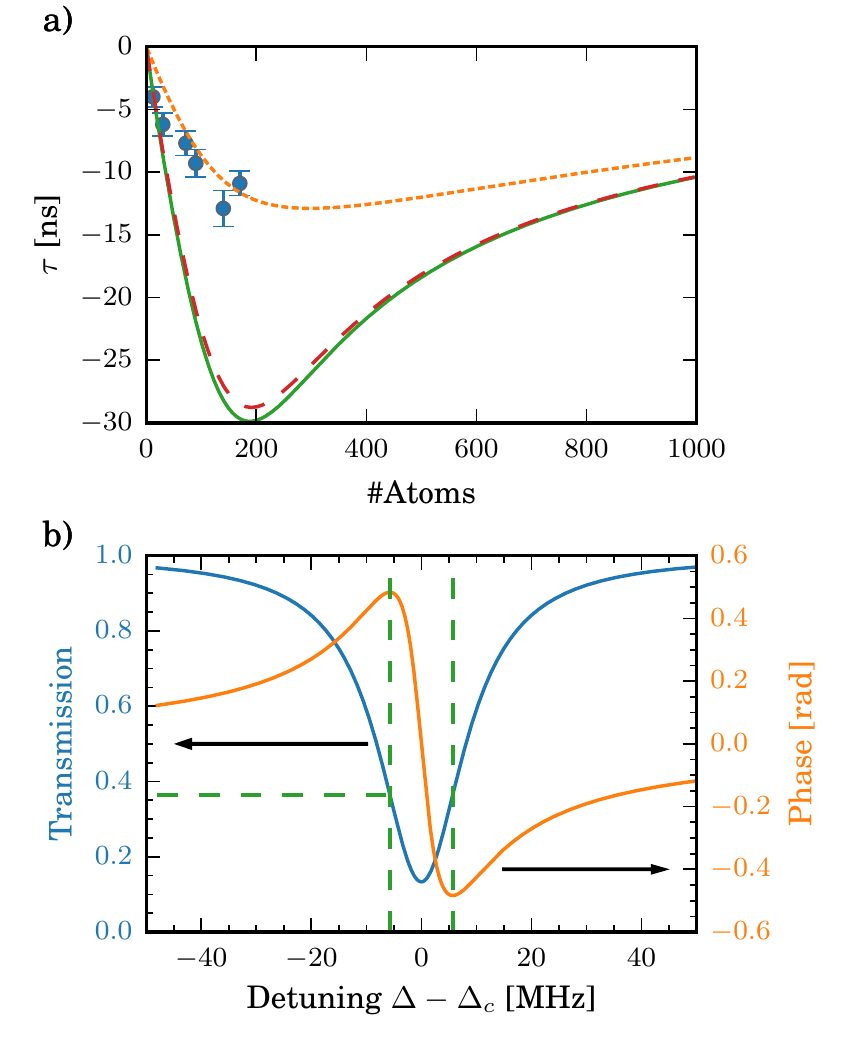}
\caption{(a) Largest group delay induced by the cold atom cloud versus number of atoms. Blue dots : measured data. Orange dotted line : simulation based on Eq.~\ref{Eq_Fouriertransform} and using the measured transfer function ${\cal S}(\omega)$ (see Eq.~\ref{Eq:S(omega)}) and incoming pulse with duration $\Delta t_0=25$\,ns (including the chirp). Green solid line : prediction $\tau_{\rm max}(N)$ based on Eq.~\ref{Eq_taumax}. Red dashed line : same as dotted line but taking a $200$\,ns long pulse (without chirp). (b) Transfer function reconstructed from our measurements in steady state for $N=180$ atoms. The horizontal axis is shifted by $\Delta_{\rm c}=-2$\,MHz. Blue line: transmission $|{\cal S}(\Delta-\Delta_{\rm c})|^2$. Orange line: phase $\phi=\arg[{\cal S}]$ imprinted on the transmitted light by the cold-atom cloud. The green dashed lines mark the extreme phase-shifts and the associated transmission level.}
\label{Fig3_summary}
\end{figure}

Figure~\ref{Fig3_summary}a shows our measurements of the largest (negative) delay achievable using a cold atomic sample with our parameters and a variable atom number. We compare the measurements with the theoretical expression
\begin{equation}\label{Eq_taumax}
\tau_{\rm max}(N)= -\frac{2}{\Gamma_{\rm c}(N)}\frac{A(N)}{1-A(N)}
\end{equation}
that is obtained by simply differentiating the phase of the transfer function ${\cal S}(\omega)$ given by Eq.~\ref{Eq:S(omega)}. Here $A(N)$ is obtained by interpolating the experimental values of the transmission $|{\cal S}(\omega)|^2$ in steady-state (see Supplemental Material). As can be seen, the measured delays are significantly smaller than the predicted ones. This is again due to the finite spectral width of the $25$-ns long pulse. The on-resonance delay calculated from the reconstruction of the transmitted pulse is on the other hand in good agreement with the data. We have cross-checked by simulating the propagation of a long pulse ($\Delta t=200$\,ns) that we indeed recover the largest theoretical delay and that the chirp would play little role in that case. Interestingly we note that the delay does not increase monotonically with the number of atoms: this comes from the saturation with $N$ of the amplitude $A(N)$ together with the increasing linewidth $\Gamma_c(N)$, which are induced by the resonant dipole-dipole interactions~\cite{Jennewein2015a}. The largest delay we would achieve, had we used a very long pulse, would be $-30$\,ns.

Finally, figure~\ref{Fig3_summary}b summarizes our measurements of the coherent optical response of the atomic cloud in steady state and for our largest atom number, $N=180$. In this case the phase shift $\phi(\omega)$ imprinted on the laser beam varies by $\sim 1$ radian over a detuning range $\sim[-\Gamma,\Gamma]$. One could thus envision using a second laser beam to induce a controlled light-shift on the atoms, and hence control the phase imprinted by the atoms on the probe laser containing less than 1 photon using that second beam.
Our system may therefore be interesting for applications in quantum optics and quantum information processing~\cite{Chang2014}.

In conclusion, we have demonstrated that a cloud of cold atoms with a size much smaller than the waist of a beam focused on it produces a large negative delay associated to a large negative group index close to an atomic resonance and yet the extinction by the cloud remains limited. We have inferred from this large delay the phase imprinted by the cloud on the laser beam and found values as large as $\sim 1$ radian. As the number of photons per pulse sent on the cloud is smaller than one, it opens interesting perspectives in quantum technologies, such as generating efficiently large non-linear phase shifts using laser beams with low-intensity, following the post-selection method demonstrated in~\cite{Feizpour2015}.

\begin{acknowledgments}
We acknowledge support from the E.U. through the ERC Starting Grant ARENA and the HAIRS project, from the Triangle de la Physique (COLISCINA project), the labex PALM (ECONOMIC project) and the Region Ile-de-France (LISCOLEM project). We thank F.~Bretenaker, F.~Goldfarb, J.~Ruostekoski, and C.S.~Adams for fruitful discussions. J.-J.~G. is a senior member of the Institut Universitaire de France.
\end{acknowledgments}

\newpage
\begin{widetext}
\section{Supplemental Material}

In this Supplemental Material, we present the full data set used for the reconstruction of the transmitted Gaussian pulse.

As explained in the manuscript, we reconstruct the transmitted pulse using Eq.~(2) of the paper and the transfer function ${\cal S}(N,\omega)$  measured independently in steady-state~\cite{Jennewein2015a}. Figure~\ref{Fig_Fitting_parameters} shows the parameters $A(N)$, $\Delta_{\rm c}(N)$ and $\Gamma_{\rm c}(N)$ in the expression of ${\cal S}(N,\omega)$, which were measured for discrete values of the atom number, and interpolated to any value of $N$ using the following functions:
\begin{eqnarray}
A(N)                &=& 1 - \sqrt{a + (1-a)\exp(-b N)}\ {\rm with} \ a=0.1\pm0.02 \ {\rm and}\  b=(1.9\pm0.2)\times10^{-2}\\
\Delta_c(N)/\Gamma  &=& a+b~N\ {\rm with}\ a=(-2.7\pm5.5)\times 10^{-2}\  {\rm and}\  b=(-1.7\pm0.5)\times10^{-3}\\
\Gamma_c(N)/\Gamma  &=& a+b~N\ {\rm with}\ a=1.44\pm0.21 \  {\rm and}\  b=(9.6\pm1.9)\times 10^{-3}\ .
\end{eqnarray}

\begin{figure*}[hbt!]
  \centering
  \includegraphics[width=8cm]{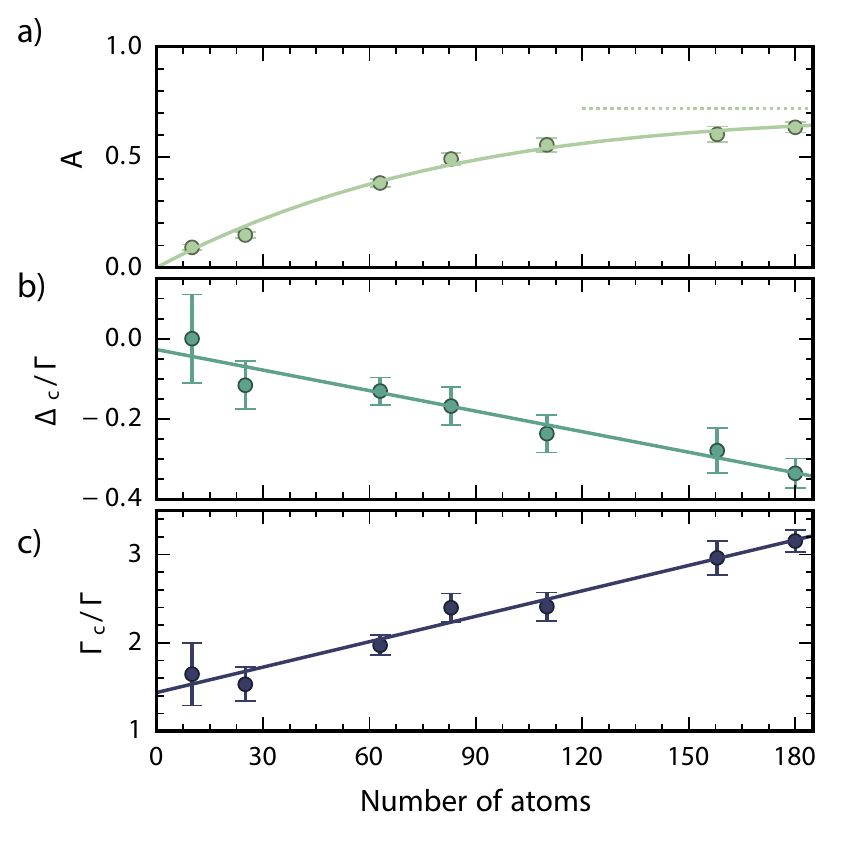}\\
  \caption{Parameters (a) $A(N)$, (b) $\Delta_{\rm c}(N)$ and (c) $\Gamma_{\rm c}(N)$, of the function ${\cal S}(N,\omega)$. Circles : data deduced from transmission spectra measured in steady state~\cite{Jennewein2015a}. Error bars are from the fits of the transmission spectra. Solid lines : fits of the data (see text). Dotted line  : asymptotic value of $A$ for $N\gg 1$.}\label{Fig_Fitting_parameters}
\end{figure*}

We then fit the reconstructed pulse using the same Gaussian function as for the data:
\begin{equation}\label{Eq_gauss}
G(t) = B \exp[-{(t-\tau)^2/ 2 \Delta t^2}]\ ,
\end{equation}
and extract $B$, $\tau$ and $\Delta t$.
We repeat this procedure for several detunings and atom numbers. Figure~\ref{Fig_reconstructed_pulse} shows the comparison between the measured amplitude, delay and width, with their reconstructed counter-parts.\\

\begin{figure*}
\centering
\subfigure[\ 12 Atoms]{
	\includegraphics[width=0.38\textwidth]{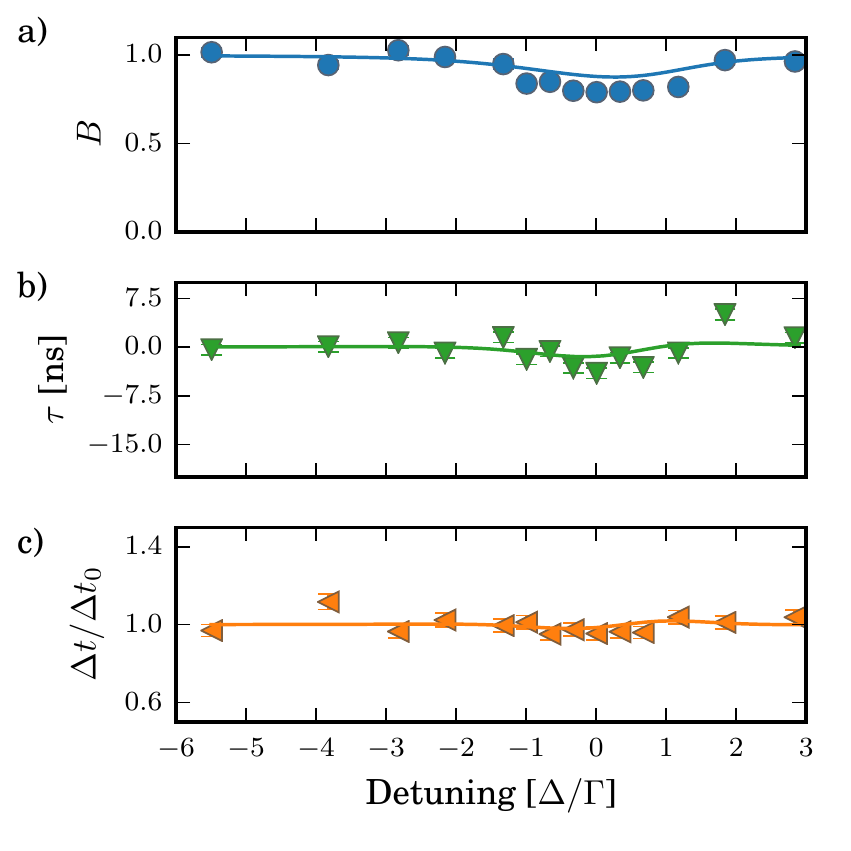}}
\subfigure[\ 30 Atoms]{
	\includegraphics[width=0.38\textwidth]{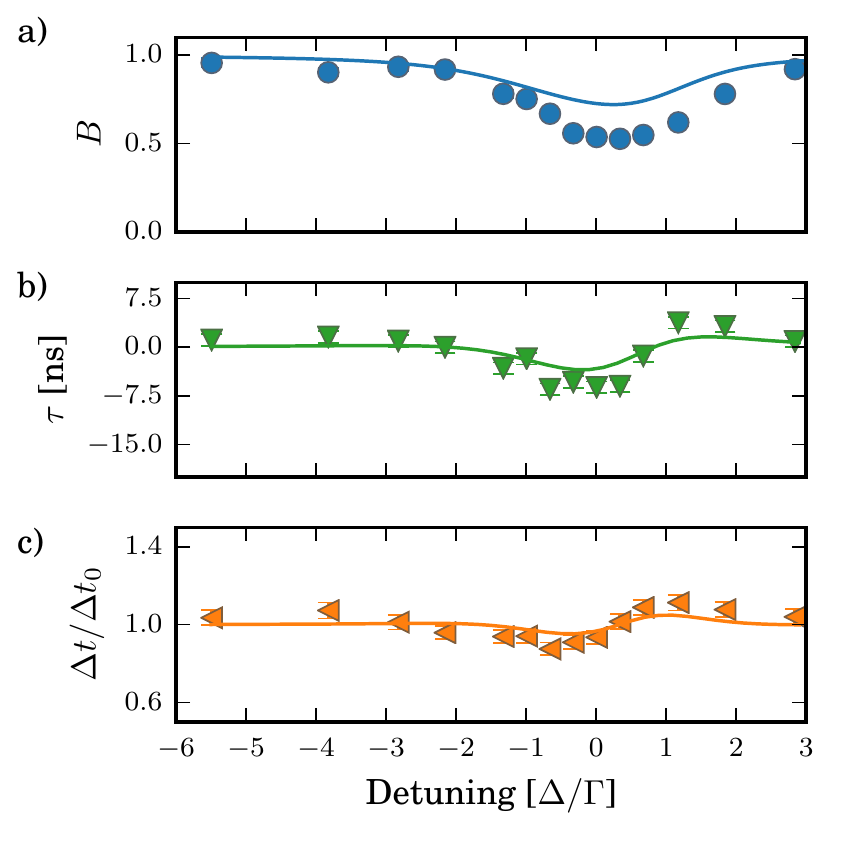}}
\subfigure[\ 72 Atoms]{
	\includegraphics[width=0.38\textwidth]{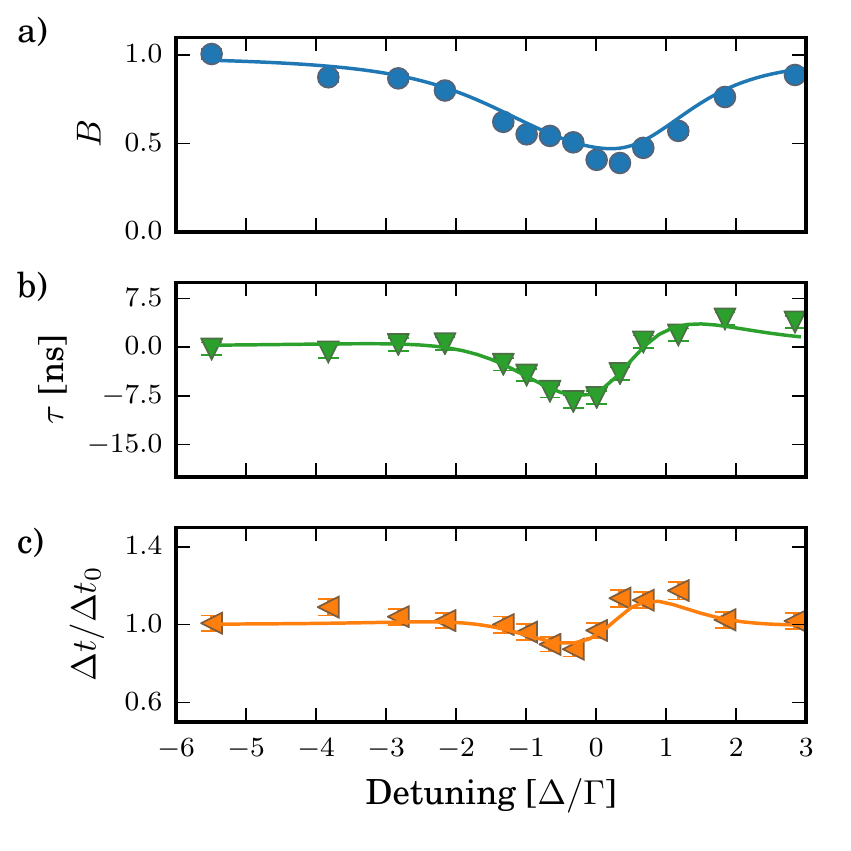}}
\subfigure[\ 90 Atoms]{
	\includegraphics[width=0.38\textwidth]{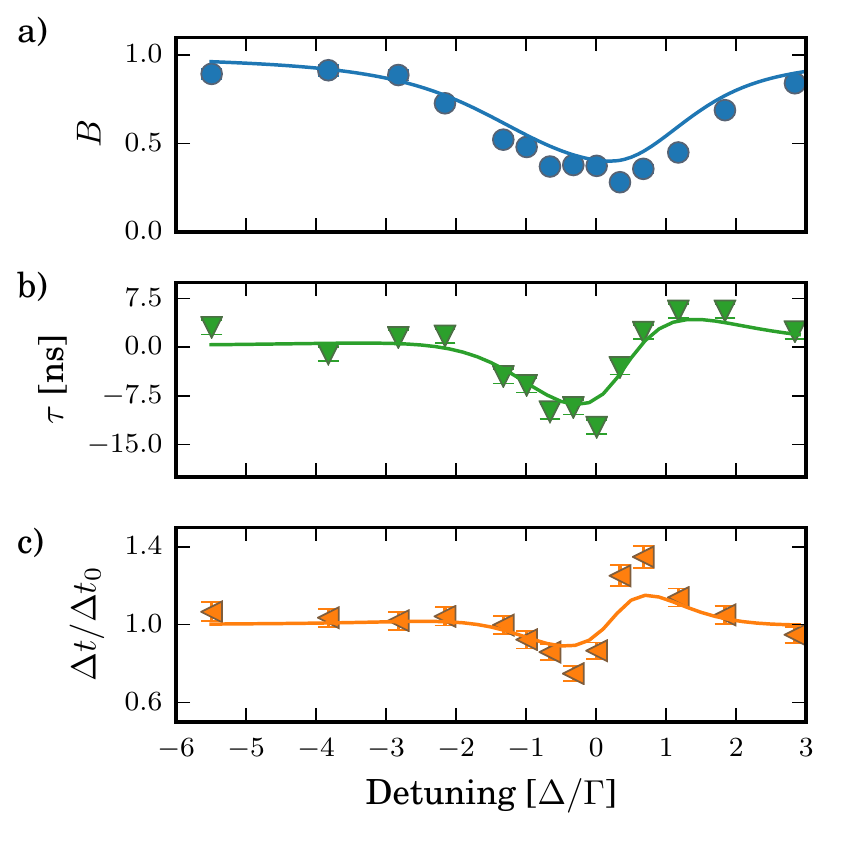}}
	\subfigure[\ 140 Atoms]{
	\includegraphics[width=0.38\textwidth]{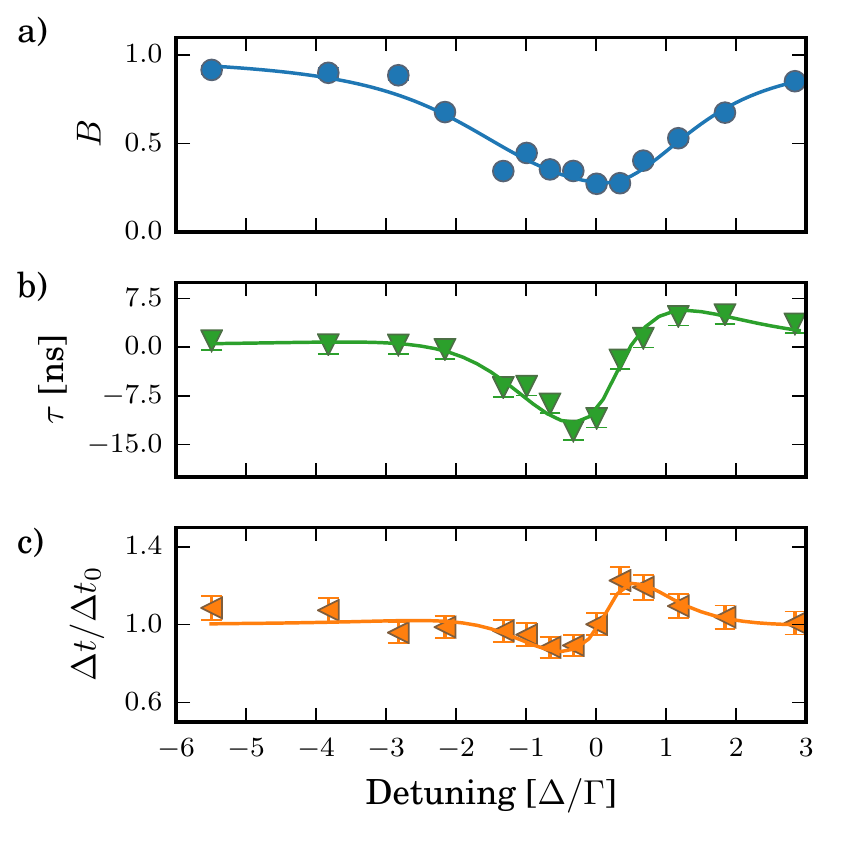}}
\subfigure[\ 170 Atoms]{
	\includegraphics[width=0.38\textwidth]{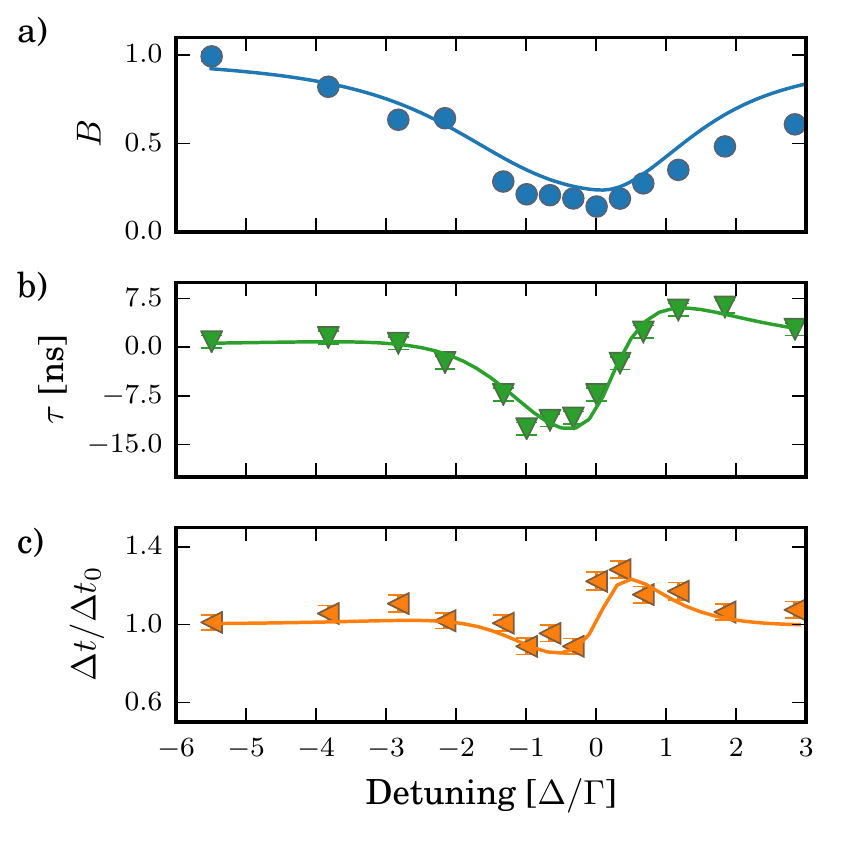}}
\caption{For each group of figures corresponding to a given number of atoms:
a) Comparison between the amplitude $B$ obtained by fitting the measured transmitted pulse (markers) and the one obtained from the fit of the reconstructed pulse (solid line).
b) Comparison between the delay $\tau$ obtained by fitting the measured pulse (markers) and the one obtained from the reconstructed pulse (solid line).
c)  Ratio between the temporal width $\Delta t$ of the pulse after transmission, and the width $\Delta t_0$ of the incident pulse measured in the absence of atoms. The markers are obtained from fits of the measured transmitted pulses. The solid lines are obtained from fits of the reconstructed pulses. Error bars are from the Gaussian fits to the data.}\label{Fig_reconstructed_pulse}
\end{figure*}

\end{widetext}

\begin{thebibliography}{80}

\bibitem{Withayachumnankul2010} W. Withayachumnankul, B.M. Fischer, B. Ferguson, B.R. Davis, D. Abbott, A systemized view of superluminal wave propagation, \doilink{10.1109/JPROC.2010.2052910}{Proc. IEEE {\bf 98}, 1 (2010)}.

\bibitem{Zangwill2012} A. Zangwill, {\it Modern Electrodynamics}, Cambridge, UK (2012).

\bibitem{Garrett1970} C.G.B. Garrett and D.E. McCumber, Propagation of a Gaussian light pulse through an anomalous dispersion medium, \doilink{10.1103/PhysRevA.1.305}{Phys. Rev. A {\bf 1}, 305 (1970)}

\bibitem{Gehring2006} G.M. Gehring, A. Schweinsberg, C. Barsi, N. Kostinski, R.W. Boyd, Observation of backward pulse propagation through a medium with a negative group velocity, \doilink{10.1126/science.1124524}{Science {\bf 312}, 895 (2006)}

\bibitem{Fleischhauer2005} M. Fleischhauer, A. Imamoglu, and J.P. Marangos, Electromagnetically induced transparency: Optics in coherent media, \doilink{10.1103/RevModPhys.77.633}{Rev. Mod. Phys. {\bf 77}, 633 (2005)}.

\bibitem{Hau1999} L.V. Hau, S.E. Harris, Z. Dutton and C.H. Behroozi, Light speed reduction to 17 metres per second in an ultracold atomic gas \doilink{10.1038/17561}{Nature {\bf 397}, 594 (1999)}.

\bibitem{Chu1982} S. Chu and S. Wong, Linear pulse propagation in an absorbing medium, \doilink{10.1103/PhysRevLett.48.738}{Phys. Rev. Lett. {\bf 48}, 738 (1982)}.

\bibitem{Tanaka2003} H. Tanaka, H. Niwa, K. Hayami, S. Furue, K. Nakayama, T. Kohmoto, M. Kunitomo, and Y. Fukuda, Propagation of optical pulses in a resonantly absorbing medium: Observation of negative velocity in Rb vapor, \doilink{10.1103/PhysRevA.68.053801}{Phys. Rev. A {\bf 68}, 053801 (2003)}.

\bibitem{Keaveney2012b} J. Keaveney, I.G. Hughes, A. Sargsyan, D. Sarkisyan, and C.S. Adams, Maximal refraction and superluminal propagation in a gaseous nanolayer, \doilink{10.1103/PhysRevLett.109.233001}{Phys. Rev. Lett. {\bf 109}, 233001 (2012)}.

\bibitem{Chang2014} D.E. Chang, V. Vuletic and M.D. Lukin, Quantum non-linear optics - photon by photon, \doilink{10.1038/nphoton.2014.192}{Nat. Phot. {\bf 8}, 685 (2014)}.

\bibitem{Jechow2013} A. Jechow, B.G. Norton, S. H\"andel, V. Blums, E.W. Streed, and D. Kielpinski, Controllable optical phase shift over one radian from a single isolated atom, \doilink{10.1103/PhysRevLett.110.113605}{Phys. Rev. Lett. {\bf 110}, 113605 (2013)}.

\bibitem{Abdullah2009} S.A. Aljunid, M.K. Tey, B. Chng, T. Liew, G. Maslennikov, V. Scarani, and C. Kurtsiefer, Phase shift of a weak coherent beam induced by a single atom, \doilink{10.1103/PhysRevLett.103.153601}{Phys. Rev. Lett. {\bf 103}, 153601 (2009)}.

\bibitem{Pototschnig2011} M. Pototschnig, Y. Chassagneux, J. Hwang, G. Zumofen, A. Renn, and V. Sandoghdar, Controlling the phase of a light beam with a single molecule, \doilink{10.1103/PhysRevLett.107.063001}{Phys. Rev. Lett. {\bf 107}, 063001(2011)}.

\bibitem{Feizpour2015} A. Feizpour, M. Hallaji, G. Dmochowski, and A.M. Steinberg, Observation of the nonlinear phase shift due to single post-selected photons, arXiv:1508.05211

\bibitem{Jennewein2015a} S.~Jennewein, M.~Besbes, N.J.~Schilder, S.D.~Jenkins, C.~Sauvan, J.~Ruostekoski, J.-J.~Greffet, Y.R.P.~Sortais, A.~Browaeys, Observation of the failure of Lorentz local field theory in the optical response of dense and cold atomic systems, arXiv:1510.08041

\bibitem{Sortais2007} Y.R.P. Sortais {\it et al.}, Diffraction-limited optics for single-atom manipulation, \doilink{10.1103/PhysRevA.75.013406}{Phys. Rev. A {\bf 75}, 013406 (2007)}.

\bibitem{Bourgain2013b} R. Bourgain, J. Pellegrino, A. Fuhrmanek, Y.R.P Sortais, and A. Browaeys, Evaporative cooling of a small number of atoms in a single-beam microscopic dipole trap, \doilink{10.1103/PhysRevA.88.023428}{Phys. Rev. A {\bf 88}, 023428 (2013)}.

\bibitem{Tey2009} M.K. Tey, G. Maslennikov, T. Liew, S.A. Aljunid, F. Huber, B. Chng, Z. Chen, V. Scarani, and C. Kurtsiefer, Interfacing light and single atoms with a lens, \doilink{doi:10.1088/1367-2630/11/4/043011}{New J. Phys. {\bf 11}, 043011 (2009)}.

\bibitem{Wigner1955} E.P. Wigner, Lower limit for the energy derivative of the scattering phase shift, \doilink{10.1103/PhysRev.98.145}{Phys. Rev. {\bf 98}, 145 (1954)}.

\bibitem{Bourgain2013a} R.~Bourgain, J.~Pellegrino, S.~Jennewein, Y.R.P~Sortais and A.~Browaeys, Direct measurement of the Wigner time delay for the scattering of light by a single atom, \doilink{10.1364/OL.38.001963}{Opt. Lett. {\bf 38}, 1963 (2013)}.

\bibitem{Sortais2012} Y.R.P.~Sortais, A.~Fuhrmanek, R.~Bourgain, and A.~Browaeys, Sub-Poissonian atom-number fluctuations using light-assisted collisions,  \doilink{10.1103/PhysRevA.85.035403}{Phys. Rev. A \textbf{85}, 035403 (2012)}.

\bibitem{Schilder2015} N.J. Schilder, C. Sauvan, J.-P. Hugonin, S. Jennewein, Y.R.P. Sortais, A. Browaeys, and J.-J. Greffet, Role of polaritonic modes on light scattering from a dense cloud of atoms, {\it submitted}, arXiv:1510.07993[physics.atom-ph].

\end{thebibliography}
\end{document}